\documentstyle[a4wide,12pt,epsfig]{article}
\newcommand{\be}{\begin{equation}}
\newcommand{\ee}{\end{equation}}

\begin{document}
\begin{flushright}
 hep-ph/mmnnyyy\\
 IISc-CTS-9/96\\
 LNF-96/018(P)
 \end{flushright}
\begin{center}
\begin{Large}
Eikonalized mini-jet cross-sections in $\gamma \gamma$ collisions\footnote{
To appear in the proceedings of the  workshop on {\it $e^+e^-$ 2000 GeV Linear
Colliders} Annecey, Gran Sasso, Hamburg (1995).}\\
%\section{\bf Eikonalized Mini-jet Cross-Sections in Photon Collisions}
\end{Large}
\vspace{1cm}
\vspace*{0.3cm}
 A. Corsetti\\ 
 INFN, Physics Department, University of Rome La Sapienza, Rome, Italy\\

R. M. Godbole\footnote{On leave of absence from Dept. of Physics, 
Uni. of Bombay, Bombay, India.} \\
Center for Theoretical Studies, Indian Institute of Sciences\\
Bangalore 560 012, India\\

and\\
G. Pancheri \\
 INFN - Laboratori Nazionali di Frascati , I00044 Frascati\\
 
\end{center} 
\vspace{1cm}
\begin{abstract}
In this note we assess the validity and uncertainties in the predictions
of the eikonalised mini-jet model for $\sigma^{inel}_{\gamma \gamma}$. 
We are able to find a choice of parameters where the predictions are
compatible with the current data. Even for this restricted range of 
parameters the predictions at the high c.m. energies, which can be reached at 
the TeV energy $e^+e^-$ colliders, differ by about $\pm 25\%$. LEP 2 data can 
help pinpoint these parameters and hence reduce the uncertainties in the 
predictions.
\end{abstract}  
\newpage
\section*{\bf Eikonalized mini-jet cross-sections}
%\section{\bf Eikonalized Mini-jet Cross-Sections in Photon Collisions}

\begin{center}
\vspace*{0.3cm}

 A.~Corsetti$^1$, R.M.~Godbole$^2$, G.~Pancheri$^3$

\vspace*{0.3cm}
\begin{tabular}{ll}
% \parbox{12.8cm}{

{\it$^1$ INFN, Univ. La Sapienza, Roma, Italy}  & \hspace*{1cm}
 {\it$^2$ CTS, IISc, Bangalore, India} \\
{\it$^3$ INFN, Frascati, Italy}

%} 
\end{tabular}
\end{center} 

In this  note we wish to assess the validity and uncertainties   of 
the eikonalized mini-jet model in predicting $\sigma_{\gamma \gamma}^{inel} $ 
and further to ascertain whether measurements at LEP-200 and HERA 
 can  constrain  various parameters of the model. In its simplest 
formulation, the eikonalized  mini--jet cross-section is given by 
\begin{equation}
\label{eikonal}
\sigma^{inel}_{ab} = P^{had}_{ab}\int d^2\vec{b}[1-e^{-n(b,s)}]
\end{equation}
where the average number of collisions at a given impact
parameter $\vec{b}$ is obtained from
\begin{equation}
\label{av_n}
n(b,s)=A_{ab} (b) (\sigma^{soft}_{ab} + {{1}\over{P^{had}_{ab}}}
\sigma^{jet}_{ab})
\end{equation} 
with 
$A_{ab} (b)$ the normalized transverse overlap of the  partons 
in the two projectiles and 
$P^{had}_{ab}$ to give the probability that both colliding particles
$a,b$ be in a hadronic state.
$\sigma^{soft}_{ab}$ is the non-perturbative part of the cross-section from 
which the factor of $P_{ab}^{had} $ has already been factored out and 
$\sigma^{jet}_{ab} $ is the hard part of the cross--section. The rise in 
$\sigma^{jet}_{ab} $ drives the rise of $\sigma_{ab}^{inel}$ with energy
\cite{cline}.  We have also assumed the factorization property
$$
P_{\gamma p}^{had} = P_{\gamma}^{had} ; 
\ \ \ P_{\gamma \gamma}^{had} =  (P_{\gamma}^{had})^2.
$$
The  predictions of the eikonalised mini-jet model 
\cite{minijets} for photon induced processes \cite{ladinsky}
depend on 1) the assumption of one or more eikonals, 2)  the hard jet 
cross-section 
$\sigma^{jet}_{ab}=\int_{p_{tmin}} {{d^2\hat{\sigma}}\over{dp_t^2}} dp_t^2$ 
which in turn depends on the minimum 
$p_t$ above which one can expect perturbative QCD to hold, viz. $ p_{tmin}$,
and the  parton densities in the colliding particles $a$ and $b$, 
3) the soft cross--section $\sigma^{soft}_{ab}$, 4) the overlap function
$ A_{ab}(b) $, defined as 
\begin{equation}
\label{aob}
A_{ab}(b)={{1}\over{(2\pi)^2}}\int d^2\vec{q}{\cal F}_a(q) {\cal F}_b(q) 
e^{i\vec{q}\cdot \vec{b}}
\end{equation}
 where ${\cal F}$ is the Fourier transform of the b-distribution
of partons in the colliding particles and 5) last but not the least 
$P_{ab}^{had}$.

In this note we shall restrict ourselves to a single eikonal. 
The hard jet cross-sections have  been evaluated  in LO perturbative QCD. 
%using two different photonic parton densities  DG \cite{DG} 
%and GRV \cite{GRVo}. 
The dependence of $\sigma_{ab}^{jet}$ on $ p_{tmin}$ 
is strongly correlated with the parton densities used. 
Here we show the results using GRV densities \cite{GRVo}
(see ref. \cite{fatlinac}  for the results using the DG densities \cite{DG}).
For the purposes of 
this note, we determine $\sigma_{\gamma \gamma}^{soft}$ 
from $\sigma_{\gamma p}^{soft}$ which is obtained by a 
fit to the photoproduction data. We use the 
Quark Parton Model  suggestion  
$\sigma_{\gamma \gamma}^{soft} = {{2}\over{3}} 
\sigma_{\gamma p}^{soft}$. 

In the original use of the eikonal model, the overlap function $A_{ab} (b)$ of
eq. (\ref{aob})  is obtained using for ${\cal F}$ the electromagnetic form 
factors and  thus, for
 photons, 
 a  number of authors 
\cite{SARC,FLETCHER} have assumed 
for ${\cal F }$   the pole expression used for the pion 
electromagnetic form factor,  on the basis of Vector Meson Dominance (VMD).
We shall investigate here another possibility, i.e. that the b-space 
distribution of partons in the photon is the Fourier transform 
of their intrinsic 
transverse momentum distributions. This will correspond to use 
the functional expression expected  for the 
perturbative part \cite{pertint} 
\begin{equation}
\label{intrinsicphot}
{{d N_{\gamma}}\over{dk_t^2}}={{1}\over{k_t^2+k_o^2}}
\end{equation}
Recently this expression was confirmed by the ZEUS \cite{ZEUS} Collaboration,
with $k_o=0.66 \pm 0.22$ GeV.
For $\gamma \gamma$ collisions, the overlap function
 is now
simply given by
\begin{equation}
\label{aobgg}
A(b)={{1}\over{4 \pi}} k_o^3 b K_1(b k_o)
\end{equation}
with $K_1$ the Bessel function of the third kind. It is interesting to notice that
 for photon-photon collisions
the overlap function will have the same analytic expression 
for both our ans\"atze: the VMD inspired pion form factor or the intrinsic
transverse momentum; the only difference being that the former corresponds to
a fixed value of $k_0 = 0.735$ GeV whereas  the latter allows us to 
vary the 
value of the parameter $k_0$.  Thus both 
 possibilities can be easily studied by simply changing $k_0$ appropriately. 
Notice that the region most important to this calculation is for large 
values of the parameter b, where the overlap function changes trend,
and is larger for smaller $k_o$ values.

As for $P^{had}_\gamma$, this is clearly expected to be ${\cal O} (\alpha_{em})$
and from  VMD one would expect $1/250$. From phenomenological considerations
\cite{FLETCHER}and fits to HERA data, one finds a value $1/200$, which 
indicates at these energies a non-VMD component of $\approx 20\%$. It should 
be noticed that 
the eikonalised minijet cross--sections do not depend  on
$A_{\gamma \gamma}$ and $P^{had}_{\gamma \gamma}$ separately, 
but depend only on the ratio of the two \cite{drees1,dgo10}.
%which for our ans\"atze for $A_{\gamma \gamma}$ means ratio of $k_0$ and 
%$P^{had}_{\gamma \gamma}$.

Having thus established the range of variability of the quantities 
involved in the calculation of
 total photonic cross sections, we now proceed to calculate and compare with
existing data the eikonalized minijet cross-section for
$\gamma \gamma$ collisions. 
We use  GRV (LO) densities and values of  $p_{tmin}$ deduced from
a best fit to photoproduction.
 As discussed in \cite{ourpaper},
it is possible to include the high 
energy points in photoproduction 
using GRV densities and $p_{tmin}=2$ GeV, but the low energy 
region would be better described  by a smaller $p_{tmin}$. This is the region 
where the rise, according to some authors, notably within the framework of 
the Dual Parton Model, is attributed to the  so-called
{\it soft Pomeron}.
%photon photon collisions,  we use $p_{tmin}=1.45$ and $2.6 GeV$ to 
%cover the full range of possibilities.
For our studies here we use $p_{tmin}=2.$ GeV.
We also use  $P_{\gamma}^{had} =1/204$
and A(b) from eq.(\ref{aobgg}) with different values 
of $k_0$. One choice for $k_0$ is the
%which correspond to one standard deviation from the ZEUS \cite{ZEUS}
%collaboration value.
pole parameter value in the photon b-distribution expression,
which includes both the intrinsic transverse momentum option
$0.66 \pm 0.22$ GeV as well as the pion form factor value, 0.735 GeV. 
The other value, 1 GeV, is a possible choice which appears to fit the
present data better than everything else.  
\begin{figure}[ht]
\begin{center}
\leavevmode
\mbox{\epsfig{file=lc2000fig.ps,width=0.7\textwidth,bbllx=30pt,bblly=70pt,bburx=570pt,bbury=750pt,angle=90}}
\end{center}
\caption{
%Total inelastic photon-photon cross-section for
%different  $p_{tmin}$ and  different
%parton b-distribution in the photon. 
Total inelastic photon-photon cross-section for
$p_{tmin}= 2.$ GeV and  different
parton b-distribution in the photon. 
The solid line corresponds to $k_0=1.$ GeV.}
% and $p_{tmin}=2.$ GeV.} 
\label{gamgam}
\end{figure}
Our predictions are shown in Fig.(\ref{gamgam}). A comparison
with existing $\gamma \gamma$ data shows that all of our 
choices are compatible with the data within the present
experimental errors. 
At high energies,
however, like the ones reachable with the 
proposed linear photon colliders,
these predictions vary by about $\pm 25 \%$. Reducing the
error in the LEP1 region and adding new data points in the c.m. region 
attainable at
LEP2, can help pinpoint and restrict the choices. Were the LEP1 and
LEP2 data to confirm  the present values, we believe that the best
representation of the present data is obtained with the higher 
$k_0$ value.
%shown in the figure with a solid line. 

\vglue0.5cm
\leftline{\bf  Acknowledgements}
\vglue0.1cm

R.M.G. wishes to acknowledge support from C.S.I.R. (India) under grant
no. \\
03(0745)/94/EMR-II. This research is supported in part by the EEC 
program ``Human Capital and Mobility", contract CT92-0026 (DG 12 COMA).

%%%%%%%%%%%%%%%%%%%the ps file for the figure called desy96fig.ps%%%%%%

\end{document}